\documentclass[12pt,journal,final,oneside,onecolumn,nofonttune]{IEEEtran}
\usepackage{mathrsfs}
\usepackage[compress]{cite}
\usepackage{hyperref}
\makeatletter
\renewcommand{\citepunct}{,\penalty\@m\hskip.13emplus.1emminus.1em}
\renewcommand{\citedash}{\hbox{--}\penalty\@m}
\makeatother

\usepackage{graphicx}
\usepackage{amsmath}
\usepackage{amssymb}
\usepackage{amsfonts}
\usepackage{psfrag}
\usepackage{float}
\usepackage{array}
\ifCLASSOPTIONcompsoc
\usepackage[tight,normalsize,sf,SF]{subfigure}
\else
\usepackage[tight,footnotesize]{subfigure}
\fi
\IEEEoverridecommandlockouts
\usepackage{setspace}
\begin{document}
\title{Tucker Decomposition For Rotated Codebook in 3D MIMO System Under Spatially Correlated Channel}
\author{
\IEEEauthorblockN{\large{Fang Yuan}}

\thanks{Copyright (c) 2015 IEEE. Personal use of this material is permitted. However, permission to use this material for any other purposes must be obtained from the IEEE by sending a request to pubs-permissions@ieee.org.

Fang Yuan is with Smart Wireless Laboratory, National Institute of
Information and Communications Technology (NICT), Japan (email:
yuanfang@nict.go.jp).}}

\maketitle \vspace{-1.5cm}
\begin{abstract}
This correspondence proposes a new rotated codebook for
three-dimensional (3D) multi-input-multi-output (MIMO) system under
spatially correlated channel. To avoid the problem of high
dimensionality led by large antenna array, the rotation matrix in
the rotated codebook is proposed to be decomposed by Tucker
decomposition into three low-dimensional units, i.e., statistical
channel direction information in horizontal and vertical directions
respectively, and statistical channel power in the joint horizontal
and vertical direction. A closed-form suboptimal solution is
provided to reduce the computational complexity in Tucker
decomposition. The proposed codebook has a significant dimension
reduction from conventional rotated codebooks, and is applicable for
3D MIMO system with arbitrary form of antenna array. Simulation
results demonstrate that the proposed codebook works very well for
various 3D MIMO systems.
\end{abstract}

\vspace{-0.5cm}
\section{Introduction}
Three-dimensional (3D) multi-input-and-multi-output (MIMO) systems
are promising to meet the ever-growing data demand in future 5th
Generation (5G) cellular networks, where large antenna arrays are
equipped at the base station (BS) to employ 3D beamforming to serve
users \cite{Nam2013,Rappaport2013}.

The promised performance of 3D MIMO system largely depends on the
accuracy of the channel direction information (CDI) obtained at the
BS. In time division duplexing (TDD) system, the CDI can be obtained
by channel estimation in uplink, but may be contaminated by pilot
reuse \cite{Marzetta2011}. In frequency division duplexing (FDD)
system, limited feedback is widely used, where the CDI is quantized
at the user and then fed back to the BS \cite{Love2003}. Yet the
feedback overhead is considered unacceptable under large antenna
array \cite{Jindal2006}.

Recently, the works in \cite{Adhikary2013,Kuo2012} have revealed
that the spatial correlation can be exploited to reduce the feedback
overhead significantly for 3D MIMO system with large antenna array.
The spatial correlation is observed very typical in MIMO channels
\cite{WINNER2}, due to the small antenna spacing in the array and
low angular spread in the propagation. It indicates that FDD is also
applicable for large antenna array system.

In limited feedback, various codebooks have been proposed for
spatially correlated channels. The Lloyd-like codebooks
\cite{Gray,Huang2011} have good quantization performance, but cannot
be used off-line. Discrete Fourier transformation (DFT) codebook is
proposed for highly correlated channels with uniform linear array
(ULA) at the BS \cite{Yang2010}, which has desirable features such
as constant modulus and finite alphabet. For 3D MIMO system with
uniform rectangular array (URA), a Kronecker-product DFT codebook
was proposed in \cite{Xie2013}. However, as shown in
\cite{Yang2010}, the performances of these pure DFT based codebooks
degrade severely when the angular spreads in the channel increase.

A well-known codebook, rotated codebook, has been proposed in
\cite{Love2006}, which transforms the codewords optimized for
uncorrelated channels (e.g., Grassmannian linear packing (GLP)
codewords) by channel correlation matrix.  The rotated codebook is
extended for multiuser MIMO system in \cite{Choi2013a}. It was
proved that the rotated codebook is asymptotically optimal in
quantizing any spatially correlated channels as the codebook size
becomes large \cite{Rao2006}. Therefore, the rotated codebook serves
as a performance upper bound for other books in practice and widely
applied. Theoretically, the rotated codebook can be applied directly
in 3D MIMO system with large antenna array. However, the high
dimensionality in the correlation matrix incurs not only heavy
feedback load for statistical information, but also high computation
complexity in the matrix operations \cite{Marzetta2011a}. Moreover,
it is challenging to acquire an accurate correlation matrix required
by the rotated codebook under high dimensionality, which is known as
the ``curse of the dimensionality'' \cite{Marimont1978}.

Many research efforts have been made for the dimension reduction in
the limited feedback. For example, the singular value decomposition
(SVD) is employed to alleviate the dimension problem by discarding
non-dominant eigenvectors in the correlation matrix \cite{Ko2009}.
Moreover, inspired by 3D MIMO system with URA, independent
quantization is considered, which naively quantizes the CDI of 3D
MIMO in horizontal and vertical directions independently and reuses
existing codebooks in each direction \cite{ALSB2011}. Independent
quantization avoids the problem of high dimensionality led by large
antenna array, while the performance degrades as the angular spreads
become large.

In this correspondence, we propose to apply the Tucker decomposition
in the rotation matrix required by the rotated codebook, which aims
at solving the problem of high dimensionality led by large antenna
array in 3D MIMO system. A closed-form solution is provided for the
Tucker decomposition, which is suboptimal but with low computational
complexity. Simulation results demonstrate that the proposed
codebook yields good performance for 3D MIMO systems with different
antenna arrays.

\emph{Notations}: $(\cdot)^T$, $(\cdot)^H$ and $(\cdot)^\ast$ are
respectively the transpose, Hermitian and conjugate operation,
$\otimes$ and $\|\cdot\|$ denotes the Kronecker product and
Frobenius norm, $\text{diagv}(\pmb{x})$ is the diagonal matrix with
diagonal entries from the vector $\pmb{x}$, and
$\text{diagm}(\pmb{X})$ is the diagonal matrix with diagonal entries
the same as the matrix $\pmb{X}$.

\section{System Model}
\subsection{Channel Model}
\begin{figure}[t]
\centering
   \includegraphics[width=3.55in]{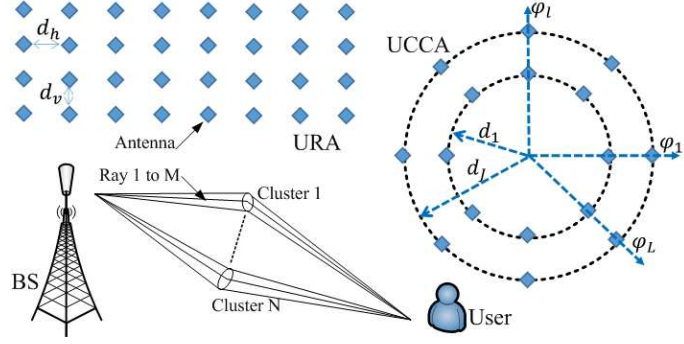}
   \caption{An example of 3D MIMO Channel, scattering clusters and antenna arrays}\label{3DMIMO1}
\end{figure}

Consider a downlink 3D MIMO system where the BS equipped with an
array of $N_t$ antennas \cite{Nam2013} to serve single-antenna users
in the macro cell, and all antennas are omni-directional. According
to the spatial channel model in \cite{Yong2005,WINNER2}, the 3D MIMO
channel from the BS to the user consists of several scattering
clusters distributed in the 3D space, as shown in Fig.
\ref{3DMIMO1}, where in each cluster there are multiple rays with
small random angle offsets. The 3D MIMO channel is expressed as a
$N_t$-dimensional vector
\begin{align}
\setlength{\abovedisplayskip}{1pt}
\setlength{\belowdisplayskip}{1pt} \pmb{h} &\triangleq
\sum\nolimits_{n=1}^N\sum\nolimits_{m=1}^M
g_{n,m}\pmb{a}_{n,m}\label{CNM}
\end{align}
where $g_{n,m}$ is the random complex gain of the $m$th ray in the
$n$th cluster, and $\pmb{a}_{n,m}$ of dimension $N_t\times1$ is the
corresponding array response.

The array response $\pmb{a}_{n,m}$ depends on the specific form of
the array mounted at the BS, as shown in \cite{Yong2005}. The array
response of URA shown in Fig. \ref{3DMIMO1} can be decomposed into
two ULA responses respectively in horizontal and vertical directions
as
\begin{align}\setlength{\abovedisplayskip}{1pt}
\setlength{\belowdisplayskip}{1pt}
\pmb{a}_{n,m}=\pmb{a}_{v}(\phi_{n,m}) \otimes
\pmb{a}_{h}(\theta_{n,m})\label{decopm}
\end{align}
with $\pmb{a}_{h}(\theta_{n,m})\!=\! [1,
e^{-jk_h{\cos\theta_{n,m}}},\!\cdots\!,
\!e^{-jk_h(N_h\!-\!1){\cos\theta_{n,m}}}]^T$,
$\pmb{a}_{v}(\phi_{n,m})\!= \![1,
e^{-jk_v{\cos\phi_{n,m}}},\!\cdots\!,\!
e^{-jk_v(N_v\!-\!1){\cos\phi_{n,m}}}]^T$, where $k_h=2\pi
\frac{d_h}{\lambda}$, $k_v=2\pi\frac{d_v}{\lambda}$,   $d_h$ and
$N_h$ ($d_v$ and $N_v$) are respectively the antenna spacing and the
number of antennas at the URA in horizontal (vertical) direction,
$\lambda$ is the carrier wavelength, $\cos\theta_{n,m}$ and
$\cos\phi_{n,m}$ are the direction cosines of the $m$th ray in the
$n$th cluster respectively in the horizontal and vertical
directions.  The array response of uniform concentric circular array
(UCCA) shown in Fig. \ref{3DMIMO1} is expressed as
\begin{align}\setlength{\abovedisplayskip}{1pt}
\setlength{\belowdisplayskip}{1pt}
\pmb{a}_{n,m}=[\pmb{a}(\varphi_{1})^T,\cdots
 \pmb{a}(\varphi_{L})^T]^T\label{decopmi}
\end{align}
with $\pmb{a}(\varphi_{l})\!=\! [e^{-j2\pi
\frac{d_1}{\lambda}{\cos(\phi_{n,m}-\varphi_{l})\cos\theta_{n,m}}},\!\cdots\!,
\!e^{-j2\pi
\frac{d_{J}}{\lambda}{\cos(\phi_{n,m}-\varphi_{l})\cos\theta_{n,m}}}]^T
$, where $J$ and $L$ are respectively as the number of rings and the
number of antennas equally placed on each ring in the UCCA,  $d_j$
is the radius of the $j$th ring, $\varphi_{l}=2l\pi/L$ is the $l$th
radial direction, $\theta_{n,m}$ and $\phi_{n,m}$ are respectively
the angle of the $m$th ray in the $n$th cluster in horizontal and
vertical directions.

\subsection{Rotated Codebook}
The CDI of 3D MIMO channel is defined as $\bar{\pmb{h}}=
{\pmb{h}}/\|\pmb{h}\|$, which is of unit-norm \cite{Love2003}. In
the procedure of limited feedback, each user quantizes the CDI using
a codebook known by the BS, and feeds the optimal codeword back to
the BS for the beamforming \cite{Love2003}. The rotated codebook is
widely applied among all the codebooks to quantize the CDI
\cite{Love2006}, and the codewords of a $B$-bit rotated codebook are
given by
\begin{align}\setlength{\abovedisplayskip}{1pt}
\setlength{\belowdisplayskip}{1pt}
\pmb{f}_i=\frac{\pmb{R}^{\frac{1}{2}}\pmb{c}_i}{\|\pmb{R}^{\frac{1}{2}}\pmb{c}_i\|},
~~i=1,\ldots,2^B\label{rote}
\end{align}
where the codewords $\pmb{c}_i$ of dimension $N_t\times 1$ are
optimized under uncorrelated channels and universal for different
users, and the rotation matrix $\pmb{R}$ is the spatial correlation
matrix in 3D MIMO channel defined as
\begin{align}\setlength{\abovedisplayskip}{1pt}
\setlength{\belowdisplayskip}{1pt}
\pmb{R}=\pmb{E}\{\pmb{h}\pmb{h}^H\},\label{frmat}
\end{align}
which differs from user to user, $\pmb{E}\{\}$ is the expectation
operation, and $\pmb{X}^{\frac{1}{2}}$ is the square root of matrix
$\pmb{X}$.

By exploiting the statistical information, the rotated codewords
$\pmb{f}_i$ in \eqref{rote} improves the quantization performance
from the universal codewords $\pmb{c}_i$ under spatially correlated
channels. It has been proved in \cite{Rao2006} that as the codebook
size increases, the rotated codebook is asymptotically optimal in
quantizing arbitrary spatially correlated channel. Yet under a large
antenna array in 3D MIMO system, the high dimensionality problem
becomes challenging. For instance, with tens or hundreds of antennas
at the array, say $N_t=64$ and $256$, the matrix $\pmb{R}$ of
dimension $N_t\times N_t$ has respectively overloaded $4096$ and
$65536$ elements.

The high dimensionality in $\pmb{R}$ not only increases the
complexity of MIMO operations, but also challenges the application
of rotated codebooks. In practice, the correlation matrix $\pmb{R}$
can be acquired at the BS by either the feedback from the user, or
the estimation from samples collected in uplink \cite{Adhikary2013}.
In high dimensionality, the former experiences a huge overhead for
each feedback of statistical information, and the latter may suffer
from the problem of ``curse of dimensionality''. As shown in
\cite{Marzetta2011a}, when the dimension of correlation matrix is
comparable with the number of samples, the widely-used
``sample-covariance'' estimation becomes invalid. Such a problem is
recognized as ``curse of dimensionality'', and is far from trivial
in the estimation theory. Therefore, we strive to reduce the
dimensionality of the rotation matrix in the rotated codebook.

\section{Tucker Decomposition for Rotated Codebook}
We propose to apply the Tucker decomposition to reduce the high
dimensionality for the correlation matrix in rotated codebook under
spatially correlated 3D MIMO channel.

\subsection{The Proposed Codeword Structure}
The proposed new rotated codebook is with the codeword structure as
\begin{align}\setlength{\abovedisplayskip}{1pt}
\setlength{\belowdisplayskip}{1pt} \pmb{f}_i =
\frac{\hat{\pmb{R}}^{\frac{1}{2}}\pmb{c}_i}{\|\hat{\pmb{R}}^{\frac{1}{2}}\pmb{c}_i\|},~~i=1,\ldots,2^B
\label{CJS2}
\end{align}
where the rotation matrix $\hat{\pmb{R}}$ is given by a designed
structure as
\begin{align}
\hat{\pmb{R}}=(\pmb{U}\!\otimes\!\pmb{V})\text{diagv}(\pmb{\lambda})
(\pmb{U}\!\otimes\!\pmb{V})^H \label{rcr}
\end{align}
which consists of three information units: the unitary matrix
$\pmb{V}$ is of dimension $N_1\times N_1$, the unitary matrix
$\pmb{U}$ is of dimension $N_2\times N_2$, and the vector
$\pmb{\lambda}$ is with nonnegative elements of dimension $N_t\times
1$ where $N_1N_2=N_t$. Both of $N_1$ and $N_2$ can be small even
when $N_t$ is large, e.g., $N_t=256$ and $N_1=N_2=16$. Thus, only
three low-dimensional information units need to be fed back from the
user to the BS for reconstructing the rotation matrix
$\hat{\pmb{R}}$, which avoids the high dimensionality problem in
conventional rotated codebooks. As will be revealed shortly, the
structure in \eqref{rcr} is led by the Tucker decomposition to the
correlation matrix $\pmb{R}$.

\subsection{Target of Each Information Unit}
The three information units in the new rotation matrix can target at
different information in the CDI of 3D MIMO channel. To see this, we
take the 3D MIMO channel with URA at the BS as an example.

Mathematically, given $N_t=N_1N_2$, any channel vector $\pmb{h}$ of
dimension $N_t\times1$ can be reshaped into a matrix form as
$\pmb{H}$ of dimension $N_1\times N_2$, e.g., via the ``reshape''
function provided in Matlab, and consequently
\begin{align}\setlength{\abovedisplayskip}{1pt}
\setlength{\belowdisplayskip}{1pt}
\pmb{H}=\text{reshape}\{\pmb{h}\}\text{~and~}
\pmb{h}=\text{vec}\{\pmb{H}\}\label{matse}
\end{align}
where $\text{vec}(\cdot)$ means vectorizing a matrix into a vector.
By setting $N_1=N_h$ and $N_2=N_v$ for URA, it has $\pmb{H}=
\sum\nolimits_{n=1}^N\sum\nolimits_{m=1}^M
g_{n,m}\pmb{a}_{h}(\theta_{n,m})\pmb{a}_{v}(\phi_{n,m})^T$, where
the columns and rows of $\pmb{H}$ stand for the channel information
in horizontal and vertical directions respectively.

The information units $\pmb{V}$ and $\pmb{U}$ can target at
statistical direction information respectively in horizontal and
vertical directions, through the SVD of the left and right
correlation matrices of 3D MIMO channel $\pmb{H}$ as
\begin{align}\setlength{\abovedisplayskip}{1pt}
\setlength{\belowdisplayskip}{1pt} \pmb{R}_h=
{\pmb{E}}\{\pmb{H}\pmb{H}^H\}=\pmb{V}\text{diagv}(\pmb{\lambda}_h)\pmb{V}^H
\text{~and~} \pmb{R}_v= {\pmb{E}}\{\pmb{H}^T
\pmb{H}^\ast\}=\pmb{U}\text{diagv}(\pmb{\lambda}_v)\pmb{U}^H\label{rigcor}
\end{align}

Let $\pmb{G}=\pmb{V}^H\pmb{H}\pmb{U}^\ast$ be the instantaneous
channel gain of $\pmb{H}$ projected onto the two unitary matrices of
$\pmb{V}$ and $\pmb{U}$, and
$\pmb{R}_g={\pmb{E}}\{\text{vec}\{\pmb{G}\}\text{vec}\{\pmb{G}\}^H\}$
be the corresponding correlation matrix with the diagonal as the
statistical channel power. Then, the information unit
$\pmb{\lambda}$ can target at the statistical channel power in the
joint horizontal and vertical direction by satisfying
\begin{align}\setlength{\abovedisplayskip}{1pt}
\setlength{\belowdisplayskip}{1pt}
{\text{diagv}(\pmb{\lambda})}={\text{diagm}(\pmb{R}_g)}\label{fr}
\end{align}
In other words, the information unit $\pmb{\lambda}$ characterizes
the interaction between the information unit $\pmb{V}$ and
$\pmb{U}$.

So far, the physical explanations for the target of each information
unit is given for URA. Actually, the explanations for horizontal and
vertical directions under URA are similar to that between transmit
and receive ends under MIMO channel studied in \cite{Raghavan2010}
and \cite{Weichselberger2006}. It should be noted that the equations
in \eqref{matse}, \eqref{rigcor} and \eqref{fr} hold for arbitrary
antenna array, not only for URA. Realizing this, the idea of
decomposing the channel information in horizontal and vertical
directions under URA can be \emph{generalized} for 3D MIMO channel
under arbitrary antenna array when \eqref{matse}, \eqref{rigcor} and
\eqref{fr} is applied. However, unlike under URA, physical
explanations for \eqref{matse}, \eqref{rigcor} and \eqref{fr} under
arbitrary antenna array may be not straightforward.

The proposed new codebook in \eqref{CJS2} is identical to the
rotated codebook in \eqref{rote} if the elements in $\pmb{G}$ are
statistically independent, where the mismatch of rotation matrices
$\|\pmb{R}-\hat{\pmb{R}}\|$ is zero, since
\begin{align}\setlength{\abovedisplayskip}{1pt}
\setlength{\belowdisplayskip}{1pt}
\pmb{R}&=\pmb{E}\{\text{vec}(\pmb{V}\pmb{G}\pmb{U}^T)\text{vec}(
\pmb{V}\pmb{G}\pmb{U}^T)^H\}=(\pmb{U}\otimes\pmb{V})\pmb{E}\{\text{vec}\{\pmb{G}\}\text{vec}\{\pmb{G}\}^H\}
(\pmb{U}\otimes\pmb{V})^H\label{keqe}\\
&=(\pmb{U}\!\otimes\!\pmb{V})\text{diagv}(\pmb{\lambda})
(\pmb{U}\!\otimes\!\pmb{V})^H=\hat{\pmb{R}}\nonumber
\end{align}
where the second equation of \eqref{keqe} is because
$\text{vec}(ABC^T)=(C\otimes A)\text{vec}(B)$. In this case, the
proposed codebook is also asymptotically optimal as the codebook
size increases like the rotated codebook in \eqref{rote}.

A sufficient condition for the gain matrix $\pmb{G}$ having
independent elements can be found similar to \cite{Raghavan2010}.
Admittedly, such independency can not be met for general channel
conditions, which leads to a nonzero mismatch of rotation matrices
$\|\pmb{R}-\hat{\pmb{R}}\|$. Then, the proposed codebook becomes
suboptimal to the rotated codebook as analyzed in \cite{Rao2006}.
Yet the performance gap between two codebooks is expected to be
small. This is because the dependency between the elements in
$\pmb{G}$ can be weakened largely after the decorrelation operation
by the SVD in \eqref{rigcor}, similar to what has been observed in
\cite{Weichselberger2006}.

\subsection{Tucker Decomposition to the Correlation Matrix $\pmb{R}$}
Note that by permutating the elements in $\pmb{h}$, the value of
$\pmb{V}$, $\pmb{U}$ and $\pmb{\lambda}$ given by \eqref{rigcor} and
\eqref{fr} is different, and so is the mismatch of rotation matrices
$\|\pmb{R}-\hat{\pmb{R}}\|$. To improve the new rotation matrix, we
consider the problem of finding the optimal $\pmb{V}$, $\pmb{U}$ and
$\pmb{\lambda}$ of given dimensionality to minimize the mismatch of
rotation matrices under arbitrary form of antenna array, which is
modeled as
\begin{align}\setlength{\abovedisplayskip}{1pt}
\setlength{\belowdisplayskip}{1pt}
\min_{\pmb{V},\pmb{U},\pmb{\lambda}} &\quad
\|\pmb{R}\!-\!(\pmb{U}\!\otimes\!\pmb{V})\text{diagv}(\pmb{\lambda})
(\pmb{U}\!\otimes\!\pmb{V})^H\|^2\label{kp}\\
\text{s.t.}& ~~\quad \pmb{V}^H\pmb{V}=\pmb{I}_{N_1},\quad
\pmb{U}^H\pmb{U}=\pmb{I}_{N_2}, \quad\pmb{\lambda}\succeq0\nonumber
\end{align}
where $\pmb{I}_{n}$ is the identity matrix of dimension $n\times n$,
and $\pmb{x}\succeq 0$ means each element in $\pmb{x}$ is no smaller
than $0$.

The problem in \eqref{kp} belongs to a classic approximation problem
of Tucker decomposition \cite{Stefan2012}, where $\pmb{\lambda}$ is
known as the core tensor, the columns in $\pmb{V}$ and $\pmb{U}$ are
respectively as tensors. Tucker decomposition is to decompose a
higher dimensional matrix into low dimensional factor matrices, and
the tensor core encompass all the possible interactions among the
low dimensional tensors in the factor matrices. Moreover, Tucker
decomposition generalizes many features in the SVD, such as
orthogonality, decorrelation and computational tractability, and
therefore is desirable for the dimension reduction in this work.

Unfortunately, the problem of Tucker decomposition in \eqref{kp} is
NP-hard in general, and thus there are few efficient algorithms in
use. Thus, we turn to find a closed-form solution for the rotation
matrix $\hat{\pmb{R}}$.

\nopagebreak{\section{Closed-form Solution for $\hat{\pmb{R}}$} To
find a closed-form solution for $\hat{\pmb{R}}$ with low complexity,
we consider to firstly optimize the $\pmb{V}$ and $\pmb{U}$ under a
structured $\pmb{\lambda}$ by exploiting the Kronecker product
decomposition. Then we optimize $\pmb{\lambda}$ by using the
obtained $\pmb{V}$ and $\pmb{U}$.
\subsection{Optimal $\pmb{V}$ and $\pmb{U}$ under a Structured
$\pmb{\lambda}$}\label{VU} First, we consider to find the optimal
$\pmb{V}$ and $\pmb{U}$ under a structured $\pmb{\lambda}$.
Specifically, we impose a Kronecker product constraint to
$\pmb{\lambda}$ as $\pmb{\lambda} =
\pmb{\lambda}_1\otimes\pmb{\lambda}_2$,  where
$\pmb{\lambda}_1\succeq0$ and $\pmb{\lambda}_2\succeq0$. Then, we
have for problem \eqref{kp}}
\begin{align}\setlength{\abovedisplayskip}{1pt}
\setlength{\belowdisplayskip}{1pt}
\!(\pmb{U}\!\otimes\!\pmb{V})\text{diagv}(\pmb{\lambda})
(\pmb{U}\!\otimes\!\pmb{V})^H=\pmb{B}\otimes\pmb{C}\label{eq19}
\end{align}
where $\pmb{B}=\pmb{U}\text{diagv}(\pmb{\lambda}_1)\pmb{U}^H$ and
$\pmb{C}=\pmb{V}\text{diagv}(\pmb{\lambda}_2)\pmb{V}^H$ are positive
semi-definite matrices.

By the structured $\pmb{\lambda}$, the problem in \eqref{kp} is
reduced to a new problem as
\begin{align}
\min_{\pmb{B},\pmb{C}} &~~ \|\pmb{R}-\pmb{B} \otimes
\pmb{C}\|^2\label{kp6}\\
\text{s.t.}& ~~\pmb{B}\in \mathbb{S}^{N_2\times N_2} \text{and}~
\pmb{C}\in \mathbb{S}^{N_1\times N_1}\nonumber
\end{align}
where $\mathbb{S}^{n\times n}$ is the space of positive
semi-definite matrices with the dimensionality of $n\times n$.

The new problem in \eqref{kp6} is known as the Kronecker product
decomposition in \cite{Nikos1997}, whose optimal solution offers a
suboptimal solution of $\pmb{V}$ and $\pmb{U}$ to the Tucker
decomposition in \eqref{kp} under a structured $\pmb{\lambda}$.
According to \cite{Nikos1997}, the optimal solution to the problem
in \eqref{kp6} is obtained by rearranging the matrix $\pmb{R}$.
Specifically, the matrix $\pmb{R}$ is divided into $N_2\times N_2$
blocks as 
\begin{align}
  \pmb{R} = \left(
              \begin{array}{ccc}
                \pmb{R}_{1,1} \!& \!\cdots \!&\! \pmb{R}_{1,N_2} \\
                \vdots \!& \!\ddots \!& \!\vdots \\
                \pmb{R}_{N_2,1} \!& \!\cdots\! & \!\pmb{R}_{N_2,N_2} \\
              \end{array}
            \right)
\end{align}
where the $(i,j)$th block denoted as $\pmb{R}_{i,j}$ is of dimension
$N_1\times N_1$. Then a rearranged matrix of dimension $N_2^2\times
N_1^2$ is generated as $
\tilde{\pmb{R}}\!=[\text{vec}(\pmb{R}_{1,1}),\text{vec}(\pmb{R}_{2,1}),\!\cdots,\!\text{vec}(\pmb{R}_{N_2,N_2})]^T.
$

Denote the largest singular value, the corresponding left and right
eigenvectors of the SVD to $\tilde{\pmb{R}}$ respectively as
$\varrho^2$, $\pmb{u}$ of dimension $N_2^2\times 1$ and $\pmb{v}$ of
dimension $N_1^2\times 1$. Then, as shown in \cite{Nikos1997}, the
optimal matrices $\pmb{B}$ of dimension $N_2\times N_2$ and
$\pmb{C}$ of dimension $N_1\times N_1$ to the problem in \eqref{kp6}
are given by
\begin{align}
\text{vec}(\pmb{B})=\varrho\pmb{u} \text{and~}
\text{vec}(\pmb{C})=\varrho\pmb{v}
\end{align}
Moreover, since $\pmb{R}$ is symmetric and positive semi-definite,
$\pmb{B}$ and $\pmb{C}$ are also symmetric and positive
semi-definite\cite{Nikos1997}, which can be regarded as the
correlation matrices. Thus, the information units $\pmb{U}$ and
$\pmb{V}$ under a structured $\pmb{\lambda}$ are obtained by the SVD
of the matrix $\pmb{B}$ and $\pmb{C}$ respectively as given after
\eqref{eq19}.

\subsection{Optimal $\pmb{\lambda}$ under the Obtained
$\pmb{U}$ and $\pmb{V}$}\label{LMD}\newpage With $\pmb{U}$ and
$\pmb{V}$ obtained in previous subsection, we further optimize the
$\pmb{\lambda}$. Observing \eqref{kp}, we find that
\begin{align}
&\quad\min_{\pmb{\lambda}}
\|\pmb{R}\!-\!(\pmb{U}\!\otimes\!\pmb{V})\text{diagv}(\pmb{\lambda})(\pmb{U}\!\otimes\!\pmb{V})^H\|^2=\min_{\pmb{\lambda}}\!\|(\pmb{U}\!\otimes\pmb{V})^H\pmb{R}(\pmb{U}\!\otimes\!\pmb{V})\!\!-\!\text{diagv}(\pmb{\lambda})\|^2\label{eq2}
\\&=\min_{\pmb{\lambda}}\!
\|\text{diagm}\left((\pmb{U}\!\otimes\pmb{V})^H\pmb{R}(\pmb{U}\!\otimes\!\pmb{V})\right)\!\!-\!\text{diagv}(\pmb{\lambda})\|^2+\|\text{off}\left((\pmb{U}\otimes\pmb{V})^H\pmb{R}(\pmb{U}\otimes\pmb{V})\right)\|^2\label{eq3}
\end{align}
where \eqref{eq2} is because the norm is unitarily invariant, and
$\text{off}(\pmb{X})$ sets the diagonal in the matrix $\pmb{X}$ into
zeros.

Therefore, the optimal $\pmb{\lambda}$ to minimize \eqref{eq3} (and
\eqref{kp}) for the given $\pmb{U}$ and $\pmb{V}$ is obtained by
choosing
\begin{align}
\text{diagv}(\pmb{\lambda}) =
\text{diagm}\left((\pmb{U}\otimes\pmb{V})^H\pmb{R}(\pmb{U}\otimes\pmb{V})\right)\label{lmdx}
\end{align}

To summarize, the rotation matrix can be constructed as in
\eqref{rcr} by using the low-dimensional information unit $\pmb{V}$,
$\pmb{U}$ obtained in subsection \ref{VU} and $\pmb{\lambda}$
obtained in subsection \ref{LMD}. The proposed solution is in
closed-form and of low complexity, which is thus desirable in
practice. The application of the proposed codebook is almost the
same as the conventional rotated codebook, differing only in the
feedback of rotation matrix. Instead of feeding back $\pmb{R}$
directly, the user firstly performs the Tucker decomposition to
$\pmb{R}$ and feeds back three obtained low-dimensional information
units $\pmb{V}$, $\pmb{U}$ and $\pmb{\lambda}$ to the BS.

\section{Simulation Results}
In this section, we evaluate the performance of proposed rotated
codebook for 3D MIMO systems with planar array by simulations. We
consider a multi-user 3D MIMO system, where the BS serves $K$
single-antenna users simultaneously with zero-forcing beamforming
and random user scheduling \cite{Jindal2006}. The average sum rate
is investigated as the performance metric, since a codebook with
better quantization performance results in a larger average sum
rate. The users have the same receive signal-to-noise ratio (SNR),
but different horizontal and vertical clusters in its own 3D MIMO
channel.

The channel spatial parameters are set similar in \cite{WINNER2}.
The horizontal and vertical angles in \eqref{CNM} are modeled
respectively as $\theta_{n,m}= \theta_{0}+\theta_{n}+\delta
\theta_{n,m}$ and $\phi_{n,m}=\phi_{0}+\phi_{n}+\delta \phi_{n,m}$,
where $\theta_{0}$ and $\phi_{0}$ are the center of horizontal and
vertical clusters uniformly distributed respectively in
$(-60^\circ,~60^\circ)$ and $(-45^\circ,~45^\circ)$, $\theta_{n}$
and $\phi_{n}$ are the horizontal and vertical deviation of the
$n$th cluster from the center, $\delta\theta_{n,m}$ and
$\delta\phi_{n,m}$ are small random offsets. We set $N=12$ and
$M=20$, model $\theta_{n}$ and $\phi_{n}$ as identically independent
distributed (i.i.d.) Gaussian variables with zero mean and a root
mean square (RMS) of $\sigma$, $\delta\theta_{n,m}$ and
$\delta\phi_{n,m}$ as Lapalacian variables with a RMS of
$1^{\circ}$, and $g_{n,m}$ are i.i.d. Gaussian variables. Two
antenna arrays are studied, i.e., URA and UCCA, where both low and
large angular spreads ($\sigma=5^{\circ}$ and $\sigma=20^{\circ}$)
are considered.

Three codebooks are evaluated in the figures: 1) A $2B$-bit rotated
codebook (labeled as ``RC'') given in \eqref{rote} quantizes the CDI
of 3D MIMO channel directly. 2) A $2B$-bit proposed codebook
(labeled as ``TDC'') given in \eqref{CJS2} quantizes the CDI of 3D
MIMO channel directly but with reduced dimension in the rotation
matrix. 3) The independent quantization (labeled as ``IQC'') given
in \cite{ALSB2011} uses two rotated codewords $\pmb{c}_h$ and
$\pmb{c}_v$ to quantize the CDI of horizonal and vertical channel
directions respectively in 3D MIMO channel, the feedback CDI at the
BS is constructed as $\pmb{c}_h\otimes\pmb{c}_v$, and the codebook
size in each direction is $B$ bits. The independent quantization
avoids the problem of high dimensionality since low-dimensional
correlation matrices $\pmb{R}_h$ and $\pmb{R}_v$ given in
\eqref{rigcor} are used as the rotation matrices in each direction.
Besides, the results with perfect CDI at the BS are provided for
comparison. In the simulations, all the correlation matrices are
perfectly fed back as in \cite{Love2006}. The codewords $\pmb{c}_i$
in different rotated codebooks are obtained from random vector
quantization, which are easy to generate but with performance close
to the GLP codebooks \cite{Jindal2006}.

\begin{figure}[t]
\centering
   \includegraphics[width=3.8in]{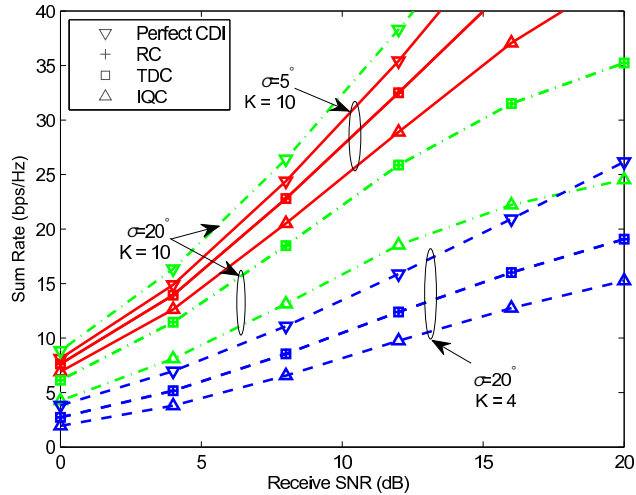}
   \caption{Average sum rates of 3D MIMO systems using URA for different angular spreads and numbers of users, where $B=8$.}\label{KandAS}
\end{figure}

Fig. \ref{KandAS} evaluates the average sum rates of three codebooks
for 3D MIMO system under URA, where $N_h=N_v=8$, $N_1=N_h$ and
$N_2=N_v$. It is shown that under different cases, the proposed
codebook in \eqref{CJS2} has a performance almost identical to the
rotated rotated codebook in \eqref{rote}. This is because URA is
particularly natural for the decomposition in \eqref{matse},
\eqref{rigcor} and \eqref{fr} and the elements in $\pmb{G}$ are
nearly independent as discussed in \cite{Raghavan2010}. Both the two
codebooks achieve much of the average sum rates obtained with the
perfect CDI at the BS. The performance gain of the proposed codebook
over the independent quantization becomes larger as the number of
users or the angular spread $\sigma$ increases. This is because that
compared to the independent quantization, the proposed codebook
considers not only the dimension reduction, but also the interaction
between horizontal and vertical directions by introducing the
information unit $\pmb{\lambda}$.

\begin{figure}[t]
\centering
   \includegraphics[width=3.8in]{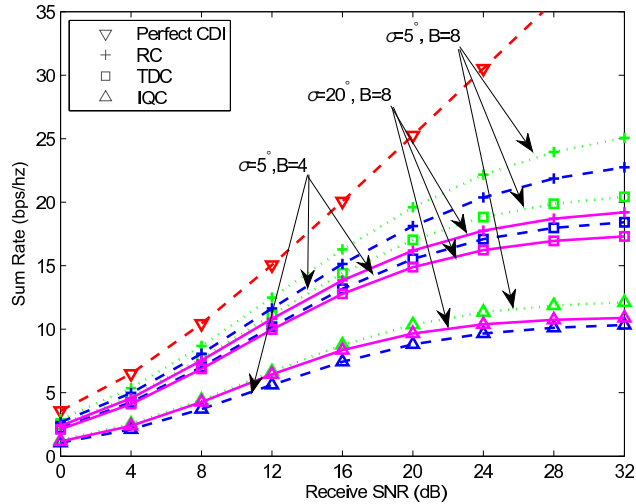}
   \caption{Average sum rates of 3D MIMO systems using UCCAs with different angle spread and codebook size,
   where $K = 4$ and $\sigma=5^\circ$.}\label{UCA}
\end{figure}

Fig. \ref{UCA} evaluates the average sum rates of three codebooks
for 3D MIMO system under UCCA, where $J=L=8$, $N_1=J$ and $N_2=L$.
The radius of UCCA is set as $d_j=0.5j\lambda$ for $j=1,\ldots,J$.
Similar simulation results can be observed for other settings. As
shown in the figure, under different angular spreads $\sigma$ and
codebook sizes, the performance of proposed codebook is slightly
inferior to that of rotated codebook. This is because under UCCA,
the mismatch of the rotation matrices $\|\pmb{R}-\hat{\pmb{R}}\|$
becomes nonzero. However, the proposed codebook reduces the problem
of high dimensionality with acceptable performance loss. Moreover,
the proposed codebook outperforms the independent quantization
significantly.

\section{Conclusions}
We have proposed a new rotated codebook for 3D MIMO system under
spatial correlated channel by introducing the Tucker decomposition.
The closed-form solution is proposed for the new rotation matrix by
Tucker decomposition with low computational complexity, and
applicable to arbitrary form of antenna arrays. Simulation results
have shown the proposed codebook works very well for 3D MIMO
systems.

\bibliographystyle{IEEEbib}
\bibliography{IEEEabrv,YF_bib}
\end{document}